\documentclass[11pt]{iopart}
\usepackage{bm}
\usepackage{hyperref}
\usepackage{amssymb}
\usepackage{amsopn}

\newcommand{\fnl}{f_{\mathrm{NL}}}
\newcommand{\tnl}{\tau_{\mathrm{NL}}}

\newcommand{\powerspectrum}{\mathcal{P}}
\newcommand{\Mfact}{\mathcal{M}}
\newcommand{\Ffact}{\mathcal{F}}
\newcommand{\Tfact}{\mathcal{T}}
\newcommand{\Kfact}{\mathcal{K}}
\newcommand{\Z}[2]{\vect{Z}_{{#1}{#2}}}
\newcommand{\ktot}{k_t}
\newcommand{\tstart}{t_{\prim}}
\newcommand{\tend}{t_{\com}}
\newcommand{\prim}{\ast}
\newcommand{\com}{\mathsf{c}}
\newcommand{\A}{\mathcal{A}}

\renewcommand{\d}{\mathrm{d}}
\newcommand{\vect}[1]{\bm{\mathrm{{#1}}}}

\begin{document}
\title{Non-gaussianity from the inflationary trispectrum}
\date{\today}
\author{David Seery and James E. Lidsey}
\address{Astronomy Unit, School of Mathematical Sciences\\
  Queen Mary, University of London\\
  Mile End Road, London E1 4NS\\
  United Kingdom}
\eads{\mailto{D.Seery@qmul.ac.uk}, \mailto{J.E.Lidsey@qmul.ac.uk}}
\submitto{JCAP}
\begin{abstract}
We present an estimate for the non-linear parameter $\tnl$, which
measures the non-gaussianity imprinted in the trispectrum of
the comoving curvature perturbation, $\zeta$. Our estimate is valid
throughout the inflationary era, until the slow-roll approximation
breaks down, and takes into account the evolution of perturbations on
superhorizon scales. We find that the non-gaussianity is always small
if the field values at the end of inflation are negligible 
when compared to their values at horizon crossing.
Under the same assumption, we show that in Nflation-type scenarios, 
where the potential is a sum of monomials, the non-gaussianity
measured by $\tnl$ is independent of the couplings and initial conditions.
\vspace{3mm}
\begin{flushleft}
  \textbf{Keywords}:
  Inflation,
  Cosmological perturbation theory,
  Physics of the early universe.
\end{flushleft}
\end{abstract}
\maketitle

\section{Introduction}
Recent observations of the cosmic microwave background (CMB) have
provided strong evidence \cite{Spergel:2006hy,Martin:2006rs,Kinney:2006qm}
that an era of inflation \cite{Starobinsky:1980te,Sato:1980yn,Guth:1980zm,
Hawking:1981fz,Albrecht:1982wi,Linde:1981mu,Linde:1983gd} occurred in the
early history of the universe. During inflation, 
the universe underwent a phase of accelerated expansion and the 
amplification of vacuum fluctuations in light fields generated 
a spectrum of density perturbations
\cite{Bardeen:1983qw,Guth:1982ec,Hawking:1982cz,Hawking:1982my}.
This spectrum is conventionally parametrized in terms of the
comoving curvature perturbation, $\zeta$,
on surfaces of uniform energy density.

A central question in modern cosmology is to determine the physics which
gave rise to inflation. It has recently become clear that a great deal of
important information is encoded in the non-linearities of the curvature
perturbation \cite{Bartolo:2004if,Rigopoulos:2004gr,Rigopoulos:2004ba,
Rigopoulos:2005xx,Allen:2005ye}.
The lower-order moments of $\zeta$
are conventionally expressed in terms of the non-linearity parameters
$\fnl$ \cite{Maldacena:2002vr,Komatsu:2001rj,Verde:1999ij}
and $\tnl$
\cite{Okamoto:2002ik,Kogo:2006kh,Boubekeur:2005fj,Lyth:2005fi,Alabidi:2005qi},
which respectively quantify the importance of the three- and
four-point correlators of $\zeta$ relative to the power spectrum.
To date, primordial non-gaussian features in the CMB have yet to be 
detected and the current upper bounds on the 
non-linearity parameters are $|\fnl| \lesssim 10^2$
\cite{Spergel:2006hy,Creminelli:2005hu,Creminelli:2006rz} and 
$|\tnl| \lesssim 10^8$ \cite{Alabidi:2005qi}. 
A major goal for the next generation of CMB experiments will be to improve
on these bounds, and the Planck
satellite is expected to be sensitive to $|\fnl| \sim 5$
\cite{Komatsu:2001rj} and $|\tnl| \sim 560$
\cite{Kogo:2006kh}. Furthermore, 
it has recently been suggested that a future all-sky survey 
of the 21-cm background has the potential to probe down to 
$|\fnl| \lesssim 0.01$ \cite{Cooray:2006km}.

Given these forthcoming improvements in the observational constraints,
there is a pressing need to obtain theoretical 
predictions for $\fnl$ and $\tnl$ in well-motivated inflationary models.
In the single-field scenario driven by a canonically normalized scalar 
field, $\zeta$ is conserved after horizon exit. This implies that
non-gaussianities cannot be generated on scales larger than the
horizon. Some non-gaussianity is imprinted by quantum mechanical
interference at the epoch of horizon crossing \cite{Lyth:2006qz}, but the
level of this effect is small. In general, one expects
$\fnl \sim \tnl \lesssim r/10$ for canonical single-field inflation
\cite{Maldacena:2002vr,Seery:2006vu}, where $r < 1$ is the tensor-to-scalar
ratio\footnote{Much larger effects are possible in models with non-canonical
scalar fields \cite{Seery:2005wm,Chen:2006nt,Huang:2006eh}, but we do not 
consider such scenarios in the present work.}.

In multiple-field models, on the other hand, 
$\zeta$ can evolve after horizon exit
due to the presence of isocurvature modes \cite{Lyth:2004gb,Malik:2005cy,
Malik:2006ir}. 
This allows additional non-gaussianity to be generated
\cite{Lyth:2005fi,Lyth:2005du,Seery:2005gb}.
This possibility has been investigated by several authors,
leading to estimates for $\fnl$
\cite{Lyth:2005qj,Zaballa:2006pv,Alabidi:2005qi,
Vernizzi:2006ve,Kim:2006te,Alabidi:2006hg,Enqvist:2004bk,
Piao:2006nm,
Gupta:2002kn,Gupta:2005nh,
Lyth:2002my,
Boubekeur:2005fj,Lyth:2006gd,Malik:2006pm,
Sasaki:2006kq,Enqvist:2005pg,
Enqvist:2004ey,Barnaby:2006cq,Battefeld:2006sz,
Huang:2005nd}
and $\tnl$
\cite{Alabidi:2005qi,Boubekeur:2005fj,Seery:2006vu,Huang:2006eh,
Sasaki:2006kq}
for a number of scenarios. In particular, Vernizzi \& Wands   
\cite{Vernizzi:2006ve} have determined the form of $\fnl$
in a two-field inflationary model with a separable potential. They find that
$|\fnl|$ is typically small in the so-called ``horizon-crossing
approximation'', where the fields roll to their minima after
inflation with values that are negligible compared to those at
horizon crossing.
The same conclusion was reached by Alabidi \& Lyth \cite{Alabidi:2005qi},
and by Kim \& Liddle \cite{Kim:2006te}
(see also Sasaki, V\"{a}liviita \& Wands \cite{Sasaki:2006kq}),
who used the horizon-crossing
approximation to compute the non-linearity parameters for multi-field
models with monomial potentials.
Recently, Battefeld \& Easther \cite{Battefeld:2006sz} have
extended the analysis of Vernizzi \& Wands to
an arbitrary number of scalar fields
and have found the corresponding form of the non-linearity parameter $\fnl$
without assuming the horizon-crossing
approximation. 

In the present paper, we derive the first complete, tree-level estimate
of the trispectrum non-linearity parameter, $\tnl$,
for inflationary models driven by multiple, uncoupled 
scalar fields.
In \S\ref{sec:nonlin} we briefly describe the non-linear $\delta N$ formalism,
and outline how it can be used to compute $\tnl$. In particular,
we present the complete, tree-level $\delta N$ expression for $\tnl$
in \S\ref{sec:fourpt}.
In \S\ref{sec:derivN}, we discuss how the derivatives of $N$ are 
computed in a scenario with an arbitrary number of 
uncoupled scalar fields and we assemble these components 
in \S\ref{sec:tnl} to derive 
an exact analytic formula for the non-linearity parameter. Our result  
is valid in the slow-roll approximation and 
accounts for the evolution of perturbations on superhorizon scales. 
In \S\ref{sec:horizoncross},
we estimate the magnitude of $\tnl$ in scenarios
driven by monomial potentials. 
We conclude with a discussion in \S\ref{sec:conclude}.

%*** Ref (1)
Throughout this paper we work in natural units where the Planck mass,
defined by $M_{\mathrm{P}}^{-2} = 8\pi G$, is set to unity.

\section{The non-linearity parameter of the trispectrum}
\label{sec:nonlin}

\subsection{The $\delta N$ formalism}

The non-linear $\delta N$ formalism provides a 
powerful method for calculating non-gaussianities in $\zeta$%
\footnote{The linear formulation was first developed by Starobinsky
\cite{Starobinsky:1986fx} and subsequently employed by Sasaki \&
Stewart \cite{Sasaki:1995aw}
to calculate the perturbation spectrum for multiple-field inflation.
It was recently generalized to the non-linear r\'{e}gime by
Lyth, Malik \& Sasaki \cite{Lyth:2004gb}. (See
also Langlois \& Vernizzi \cite{Langlois:2006vv},
and Rigopoulos \& Shellard 
\cite{Rigopoulos:2004gr} for alternative, but 
equivalent, formalisms).}.
It is valid under the assumption that causally disconnected regions
of the universe evolve locally like an unperturbed universe
in which the pressure and density are allowed to vary
\cite{Wands:2000dp,Rigopoulos:2003ak,Salopek:1990jq}.
This leads to the relation $\zeta = \delta N$,
where $N(\phi,\rho)$ represents the number of e-folds of expansion
between some flat primordial slice, $\prim$, at time
$\tstart$ on which the fields $\{ \phi_\alpha \}$ have assigned values,
and a final uniform-density surface, $\com$, at time $\tend$ 
on which the energy density has the specified value $\rho$ 
\cite{Starobinsky:1986fx,Sasaki:1995aw,Lyth:2004gb,Lyth:2005du,Lyth:2005fi}.

If slow-roll inflation is valid on the initial
flat slice, the curvature perturbation can be written as a power series
in the field perturbations evaluated at that time \cite{Lyth:2005du}:
\begin{equation}
  \label{zeta}
  \zeta = \sum_{\alpha} N_{,\alpha \prim} \delta \phi_{\alpha \prim} +
          \frac{1}{2} \sum_{\alpha \beta} N_{,\alpha\beta \prim}
          \delta \phi_{\alpha \prim} \delta \phi_{\beta \prim} + \cdots ,
\end{equation}
where $N_{,\alpha \prim} = \partial N/\partial \phi_{\alpha \prim}$,
$N_{,\alpha \beta \prim} = \partial^2 N/\partial \phi_{\alpha \prim}
\partial \phi_{\beta \prim}$, and a subscript $\prim$ denotes evaluation
at $\tstart$. Note that although $N_{,\alpha \prim}$ is
a derivative with respect to the fields on $\prim$, it depends
on the values of the fields on \emph{both} $\prim$ and $\com$. 
This is also true for the higher derivatives in Eq.~\eref{zeta}. 

We consider inflation where $M$ scalar fields contribute to the 
energy density of the universe through a separable 
potential\footnote{In this paper we
will not use the summation convention on scalar field indices 
$\{ \alpha, \beta, \gamma, \ldots \}$.}  
\begin{equation}
  \label{defW}
  W(\phi)
  = \sum_\alpha V_\alpha(\phi_\alpha)  .
\end{equation}
The scalar field dynamics can be determined in terms of the 
slow-roll parameters
\begin{eqnarray}
  \label{epsilonsro}
    \epsilon_\alpha \equiv \frac{1}{2} \left( \frac{W_{,\alpha}}{W} \right)^2 ,
    \\
  \label{etasro}
    \eta_{\alpha} \equiv \frac{W_{, \alpha \alpha}}{W} ,
    \\
  \label{xisro}
    \xi_{\alpha} \equiv \frac{1}{2} 
\frac{W_{,\alpha} W_{,\alpha\alpha\alpha}}{W^2} .
\end{eqnarray}
% *** Ref (2), Ref (3)
The slow-roll approximation is valid when the $\epsilon_\alpha$
and $|\eta_{\alpha}| \ll 1$. Another important quantity is
$\epsilon \equiv - \dot{H}/H^2$, which measures how far the inflationary
era lies from de Sitter space, with $\epsilon = 0$ for exact de Sitter
and $\epsilon < 1$ for inflation.
When the slow-roll approximation applies, one can express
$\epsilon$ in terms of the individual $\epsilon_{\alpha}$ such that
\begin{equation}
  \label{defepsilon}
  \epsilon = \sum_\alpha \epsilon_\alpha .
\end{equation}
In this r\'{e}gime the equations of motion for the scalar fields reduce to
\begin{equation}
  3H \dot{\phi}_\alpha = - W_{,\alpha} .
  \label{fieldeq}
\end{equation}
Since these are first-order equations, the phase space of the system
is $M$-dimensional.

Finally, the number of e-folds of expansion $N$,
whose derivatives appear in Eq.~\eref{zeta}, can be written as a quadrature:
\begin{equation}
  N = \int_{\tstart}^{\tend} H \, \d t
  = - \int_{\prim}^{\com} \sum_\alpha \frac{V_\alpha}{W_{,\alpha}} \;
  \d \phi_\alpha .
  \label{efolds}
\end{equation}

\subsection{The four-point correlator}
\label{sec:fourpt}

The power spectrum of $\zeta$ is formed by combining 
two copies of Eq.~\eref{zeta} and
taking the expectation value in the quantum vacuum state. 
If only the lowest derivatives of $N$ are retained, it follows 
that  
\begin{equation}
  \langle \zeta(\vect{k}_1) \zeta(\vect{k}_2) \rangle
  = (2\pi)^3 \delta(\vect{k}_1 + \vect{k}_2)
  P_{\prim}(k_1)
  \sum_{\alpha} N_{,\alpha \prim}^2 ,
\end{equation}
where $P_{\prim}(k_1)$ is the
power spectrum of a massless scalar field evaluated on $\prim$ 
and is related to the ``dimensionless'' power spectrum 
$\powerspectrum_{\prim}$ by 
\begin{equation}
  \powerspectrum_{\prim}(k) = \frac{k^3 P_{\prim}(k)}{2\pi^2} =
  \frac{H_{\prim}^2}{4\pi^2} .
\end{equation}

The four-point correlator is derived in a similar fashion by forming
a product of four copies of Eq.~\eref{zeta} and taking expectation values.
In general, the trispectrum 
is defined by $\langle \zeta(\vect{k}_1) \zeta(\vect{k}_2) \zeta(\vect{k}_3)
\zeta(\vect{k}_4) \rangle = (2\pi)^3 \delta(\sum_i \vect{k}_i)
T(\vect{k}_1,\vect{k}_2,\vect{k}_3,\vect{k}_4)$, where only 
the connected part of the correlator is considered
\cite{Okamoto:2002ik,Kogo:2006kh}. Observational 
constraints on the trispectrum are conventionally given in terms of a 
non-linearity parameter, $\tnl$, which is defined by
\cite{Boubekeur:2005fj,Lyth:2005fi,Alabidi:2005qi}
\begin{equation}
  \label{tnl}
  \fl
  T(\vect{k}_1,\vect{k}_2,\vect{k}_3,\vect{k}_4)
  = \frac{1}{2}\tnl \Big( \sum_\alpha N_{,\alpha \prim}^2 \Big)^3
  [ P_{\prim}(k_1) P_{\prim}(k_2)
  P_{\prim}(k_{14}) + \mbox{23 permutations} ] ,
\end{equation}
where $\vect{k}_{ij} = \vect{k}_i + \vect{k}_j$.
In general, $\tnl$ is a function of the four momenta
$\{ \vect{k}_1, \vect{k}_2, \vect{k}_3, \vect{k}_4 \}$
which parametrize the trispectrum.
It is convenient to rewrite Eq.~\eref{tnl} in such a way 
that the momentum-dependence is made explicit.
If we define a function $\Tfact$ such that 
\begin{equation}
  \Tfact \equiv \sum_{i < j} \sum_{m \neq i, j} k_i^3 k_j^3(k_{im}^{-3} +
  k_{jm}^{-3}) ,
\end{equation}
we find that \cite{Seery:2006vu}
\begin{equation}
  T = \frac{4 \pi^6}{\prod_i k_i^3} \tnl
  \Big( \sum_{\alpha} N_{,\alpha \prim}^2 \Big)^3
  \powerspectrum_{\prim}^3 \Tfact .
\end{equation}

After taking the product of four copies of Eq.~\eref{zeta} and evaluating
the resulting convolutions, it can be shown that there are 
four distinct terms which contribute to
$\tnl$ and in which unconstrained integrals over momenta do not appear
\cite{Seery:2006vu}. We will refer to such terms as the {\em tree-level}
terms. The leading tree-level term was computed explicitly in
Ref.~\cite{Seery:2006vu} and found to have the form 
\begin{equation}
  \Delta\tnl^{(1)} = \frac{\Ffact_4}{\Tfact
  \left( \sum_{\alpha} N_{,\alpha \prim}^2 \right)} ,
  \label{tnla}
\end{equation}
where $\Ffact_4 = \sum_{\mathrm{perms}} \Mfact_4(\vect{k}_1,
\vect{k}_2,\vect{k}_3,\vect{k}_4)$, and $\Mfact_4$ is a momentum-dependent
form-factor given by \cite{Seery:2006vu}
% ***
\begin{eqnarray}
  \fl\nonumber
  \Mfact_4 = - 2 \frac{k_1^2 k_3^2}{k_{12}^2 k_{34}^2} \frac{W_{24}}{k_t}\left[
      \frac{\Z{1}{2}\cdot\Z{3}{4}}{k_{34}^2}
      + 2 \vect{k}_2 \cdot \Z{3}{4}
      + \frac{3}{4} \sigma_{12} \sigma_{34} \right]
    \\ \mbox{} -
    \frac{1}{2} \frac{k_3^2}{k_{34}^2} \sigma_{34} \left[
      \frac{\vect{k}_1 \cdot \vect{k}_2}{k_t} W_{124} +
      \frac{k_1^2 k_2^2}{k_t^3} \left( 2 + 6 \frac{k_4}{k_t} \right)
    \right] .
\end{eqnarray}
The three momentum-dependent scalars $\sigma_{ij}$, $W_{ij}$ and
$W_{\ell m n}$ are defined by
\begin{equation}
  \sigma_{ij} \equiv \vect{k}_i \cdot \vect{k}_j + k_j^2 ,
\end{equation} 
\begin{equation}
  W_{ij} \equiv 1 + \frac{k_i + k_j}{k_t} +
  \frac{2 k_i k_j}{k_t^2} ,
\end{equation}
\begin{equation}
  W_{\ell m n} \equiv 1 + \frac{k_{\ell} + k_m + k_n}{k_t} +
  \frac{2(k_{\ell} k_m + k_{\ell} k_n + k_m k_n)}{k_t^2} +
  \frac{6 k_{\ell} k_m k_n}{k_t^3} ,
\end{equation}
and $\Z{i}{j}$ satisfies
$\Z{i}{j} \equiv \sigma_{ij} \vect{k}_i - \sigma_{ji} \vect{k}_j$.
The total
scalar momentum is written $k_t = k_1 + k_2 + k_3 + k_4$.

It was further shown in Ref.~\cite{Seery:2006vu}
that $|\Delta\tnl^{(1)}| \lesssim r/50$, where $r < 1$ is the
tensor-to-scalar ratio. This term is therefore small.

The remaining three contributions were also presented 
in Ref.~\cite{Seery:2006vu} for single-field inflation. 
For a multiple-field scenario, they are given by 
\begin{equation}
  \Delta \tnl^{(2)} =
  \frac{\sum_{\alpha\beta} \dot{\phi}_{\beta \prim} N_{,\alpha\beta\prim}
  N_{,\alpha \prim}}{\left( \sum_{\alpha} N_{,\alpha \prim}^2 \right)^2}
  \frac{\Kfact}{4 H_{\prim}} ,
  \label{tnlb}
\end{equation}
\begin{equation}
  \Delta \tnl^{(3)} = \frac{\sum_{\alpha\beta\gamma}
  N_{,\alpha\beta \prim} N_{,\alpha\gamma \prim} N_{,\beta \prim}
  N_{,\gamma \prim}}
  {\left( \sum_{\alpha} N_{,\alpha \prim}^2 \right)^3} ,
  \label{tnlc}
\end{equation}
and 
\begin{equation}
  \Delta \tnl^{(4)} = 2
  \frac{\sum_{\alpha\beta\gamma} N_{,\alpha\beta\gamma\prim}
  N_{,\alpha\prim} N_{,\beta\prim} N_{,\gamma\prim}}
  {\left( \sum_{\alpha} N_{,\alpha \prim}^2 \right)^3}
  \frac{\sum_i k_i^3}{\Tfact} ,
  \label{tnld}
\end{equation}
respectively, where 
$\Kfact$ is defined by
\begin{equation}
  \Kfact \equiv  \frac{1}{\Tfact} \sum_{\mathrm{perms}}
  \frac{k_1^3}{k_{12}^3} \Mfact_3(k_{12},k_3,k_4) 
\end{equation}
and 
\begin{equation}
  \Mfact_3(k_1,k_1,k_2) \equiv  - k_1 k_2^2 -
  4 \frac{k_2^2 k_3^2}{\ktot} + \frac{1}{2} k_1^3 +
  \frac{k_2^2 k_3^2}{\ktot^2}(k_2 - k_3) .
\end{equation}
Eq.~\eref{tnlc} was derived previously by Alabidi \& Lyth
\cite{Alabidi:2005qi} and Lyth
\cite{Lyth:2006gd}. An expression equivalent to
Eq.~\eref{tnld} was given by
Sasaki, V\"{a}liviita \& Wands \cite{Sasaki:2006kq}
in the curvaton scenario.

In general, there are an infinite number of other contributions to
$\tnl$ which involve `loops' in the sense of unconstrained integrals over
momenta \cite{Zaballa:2006pv,Lyth:2006gd}.
Such loops are suppressed relative to the tree-level
by powers of the spectrum, $\powerspectrum_{\prim} < 10^{-10}$.
In principle, however, these loop integrals may be sufficiently large 
for such terms to become important and it is not known whether 
they are always negligible in comparison with the tree-level
contributions. On the other hand,  it was recently shown
by Sloth that the contribution from a
specific one-loop correction to the power spectrum
is only about 1\% that of the tree-level effect
\cite{Sloth:2006az}. In view of this, we will 
assume that the contribution of these corrections is indeed negligible.

\section{The derivative of $N$}
\label{sec:derivN}

Our aim is to derive expressions for the 
contributions~\eref{tnla}--\eref{tnld} in terms of 
the parameters~\eref{epsilonsro}--\eref{xisro}
for a general multi-field, slow-roll inflationary scenario
when the potential is separable.
The central problem is to compute the derivatives of $N$
which appear in Eq.~\eref{zeta}.
In the horizon-crossing approximation, one ignores any contribution
from $\com$, in which case the derivatives can be
written entirely in terms of quantities evaluated on $\prim$
\cite{Lyth:1998xn,Lyth:2005qj,Alabidi:2005qi,Kim:2006te}.
The method of Vernizzi \& Wands \cite{Vernizzi:2006ve}
evades this assumption and 
was recently extended to an arbitrary number of uncoupled fields 
by Battefeld \& Easther \cite{Battefeld:2006sz}. An alternative 
approach was followed by Lyth \& Rodr\'{\i}guez
\cite{Lyth:2005qj} and by Alabidi
\cite{Alabidi:2006hg}, but in this latter case 
it is necessary to explicitly
integrate Eq.~\eref{fieldeq} between $\prim$ and $\com$.

To proceed, let us consider a point in phase space that is 
specified by the values of the fields $\{ \phi_{\alpha} \}$.
This point lies on a particular trajectory. However, 
changing the values of the fields will move the point and 
consequently will also alter the trajectory. We need 
to compute the variation of $N$ under this
change of trajectories. 
It follows from differentiation of Eq.~\eref{efolds} that
if the field values on $\prim$ are changed by $\{ 
\d \phi_{\alpha \prim} \}$, the total differential of $N$ is given by 
\begin{equation}
  \d N = \sum_\alpha \left( \left.\frac{V_\alpha}{W_{,\alpha}}
\right|_{\prim}
  - \sum_\beta \left. \frac{V_\beta}{W_{, \beta}}\right|_{\com}
  \frac{\partial \phi_{\beta \com}}{\partial \phi_{\alpha \prim}} \right)
  \; \d \phi_{\alpha \prim} ,
  \label{N-derivatives}
\end{equation}
where subscripts $\prim$ and $\com$ denote quantities evaluated 
on the initial flat hypersurface and the final comoving
hypersurface, respectively. 

The values $\phi_{\alpha \com}$ are completely determined by the trajectory on
which they lie. Since we are assuming that slow-roll inflation  
is still occurring at the final comoving hypersurface, the energy 
density $\rho$ on $\com$ is dominated by $W_{\com}$, which is constant 
over the entire surface. Using this property, 
Battefeld and Easther show that the derivatives of $\phi_{\alpha \com}$
with respect to $\phi_{\beta \prim}$ are given by
\cite{Battefeld:2006sz}\footnote{This formula was derived
in Ref.~\cite{Battefeld:2006sz} given
a specific choice of conserved quantities which label the allowed
inflationary trajectories. In general, one can always rearrange the
conserved quantities by taking arbitrary linear combinations and we have
verified that Eq.~\eref{field-derivatives} applies for this more general
choice.}
\begin{equation}
\label{field-derivatives}
  \frac{\partial \phi_{\alpha \com}}{\partial \phi_{\beta \prim}}
  =- \frac{W_{\com}}{W_{\prim}} 
  \sqrt{\frac{\epsilon_{\alpha \com}}{\epsilon_{\beta \prim}}} \left(
  \frac{\epsilon_{\beta \com}}{\epsilon_{\com}} - \delta_{\alpha \beta}
  \right)  .
\end{equation}

In obtaining the derivatives of $N$, it will prove convenient 
to introduce another set of parameters,
\begin{equation}
  E_{\alpha\beta} = \frac{\epsilon_{\alpha \com}}{\epsilon_{\com}}
  - \delta_{\alpha\beta}   ,
  \label{E}
\end{equation}
\begin{equation}
  Y_{\gamma} = \epsilon_{\gamma \com} \left(
  1 - \frac{\eta_{\gamma \com}}{\epsilon_{\com}}
  \right) ,
  \label{Y}
\end{equation}
and 
\begin{equation}
\label{X}
X_{\alpha} = \epsilon_{\alpha \com} \left[ \eta_{\alpha \com}
  \left(1 - \frac{\eta_{\alpha \com}}{\epsilon_{\com}} \right)
  - \frac{\xi_{\alpha \com}}{\epsilon_{\com}} \right] ,
\end{equation}
which are defined in terms of the values of the slow-roll parameters
on $\com$. We will also need expressions for
the variations of the slow-roll parameters on $\com$ with respect to a field
on $\prim$. In particular, we will make use of the following
relationships:
\begin{equation}
  \frac{\partial \epsilon_{\alpha \com}}{\partial\phi_{\lambda \prim}} =
  \frac{\sqrt{2}}{\sqrt{\epsilon_{\lambda \prim}}}
  \frac{W_{\com}}{W_{\prim}}
  \epsilon_{\com} (Y_\alpha - \epsilon_{\alpha \com}) E_{\lambda \alpha} ,
  \label{epsa}
\end{equation}
\begin{equation}
  \frac{\partial \epsilon_{\com}}{\partial\phi_{\lambda \prim}} =
  \frac{\sqrt{2}}{\sqrt{\epsilon_{\lambda \prim}}}
  \frac{W_{\com}}{W_{\prim}}
  \epsilon_{\com}
  \sum_{\gamma} E_{\lambda \gamma} Y_{\gamma}  ,
  \label{epsb}
\end{equation}
and
\begin{equation}
  \frac{\partial \eta_{\alpha \com}}{\partial \phi_{\lambda \prim}} =
  - \frac{\sqrt{2}}{\sqrt{\epsilon_{\lambda \prim}}}
  \frac{W_{\com}}{W_{\prim}}
  \xi_{\alpha \com} E_{\lambda \alpha} .
  \label{etaa}
\end{equation}
These follow after differentiation of Eqs.~\eref{epsilonsro}--\eref{etasro} 
and substitution of  Eqs.~\eref{field-derivatives}--\eref{X}. 
Finally, taking the derivatives of Eqs.~\eref{E} and~\eref{Y} 
with respect to $\phi_{\mu \prim}$ yields 
\begin{equation}
\label{Ealphabetaderiv}
\frac{\partial E_{\alpha\beta}}{\partial \phi_{\mu \prim}}
=- \frac{\sqrt{2}}{\sqrt{\epsilon_{\mu \prim}}} 
\frac{W_{\com}}{W_{\prim}} \left( 
\epsilon_{\alpha \com} E_{\mu\alpha} +\sum_{\gamma}
Y_{\gamma} E_{\alpha\gamma}E_{\mu\gamma} \right)  ,
\end{equation}
\begin{equation}
\label{Yalphaderiv}
\frac{\partial Y_{\gamma}}{\partial \phi_{\mu \prim}} =
- \frac{\sqrt{2}}{\sqrt{\epsilon_{\mu \prim}}} 
\frac{W_{\com}}{W_{\prim}} \left( X_{\gamma} E_{\mu\gamma}
+\left( Y_{\gamma}-\epsilon_{\gamma \com} \right) 
\sum_{\delta} E_{\mu\delta} Y_{\delta} \right)  ,
\end{equation}
where we have employed Eqs.~\eref{X}--\eref{etaa}.  

The first two derivatives of $N$ were calculated in Refs. 
\cite{Vernizzi:2006ve,Battefeld:2006sz} in terms of a parameter,
$Z_{\alpha \com}$, defined by
\begin{equation}
  Z_{\alpha \com} \equiv \frac{\epsilon_{\alpha \com}}{\epsilon_{\com}}
  W_{\com} - V_{\alpha \com} .
  \label{Z}
\end{equation}
This quantity is evaluated purely on the final comoving hypersurface $\com$
and measures the importance of any isocurvature effects 
in the generation of non-gaussian statistics. It vanishes identically  
in single-field inflation and, more generally, $\zeta$ is conserved on 
super-horizon scales whenever $Z_{\alpha \com}=0$. In this case,  
the derivatives of $N$ are controlled entirely by the potentials 
and the slow-roll parameters
evaluated on $\prim$. On the other hand, 
when more than one scalar field contributes to the energy density 
during inflation, $\zeta$ can evolve on large scales  
due to the contribution from $Z_{\alpha \com}$.

By employing Eqs.~\eref{N-derivatives} and~\eref{field-derivatives}, it 
can be shown that the derivative of $N$ with respect to a field
value on $\prim$ is given by 
\begin{equation}
  N_{,\alpha \prim} = \frac{u_{\alpha}}{\sqrt{2 \epsilon_{\alpha \prim}}}
  ,
  \label{None}
\end{equation}
where 
\begin{equation}
  u_{\alpha} \equiv    \frac{V_{\alpha \prim} + Z_{\alpha \com}}{W_{\prim}} .
\end{equation}
The second derivative of $N$ is then derived by differentiating
Eq.~\eref{None}: 
\begin{equation}
  N_{,\alpha\beta \prim} =  \delta_{\alpha\beta}
  \left( 1 - \frac{\eta_{\alpha \prim}}
  {2\epsilon_{\alpha \prim}} u_{\alpha}
  \right) +
  \frac{1}{W_{\prim} \sqrt{2 \epsilon_{\alpha \prim}}}
  \frac{\partial Z_{\alpha \com}}{\partial \phi_{\beta \prim}} .
  \label{Ntwo}
\end{equation}
It follows that $N$ only has a non-trivial, mixed
second derivative if there is a non-negligible
isocurvature contribution.

To compute the non-linearity parameter $\tnl$, we require the Taylor expansion
of $N$ up to and including the third derivative. We 
find after some algebra and use of Eqs.~\eref{E} 
and~\eref{Yalphaderiv} that differentiation of Eq.~\eref{Ntwo} yields 
\begin{eqnarray}
  \fl\nonumber
  N_{,\alpha\beta\gamma \prim} = \delta_{\alpha\beta\gamma}
  \left( \frac{\eta_{\alpha \prim}^2 -
  \xi_{\alpha \prim}}{\sqrt{2\epsilon_{\alpha \prim}^3}}
  u_{\alpha} -
  \frac{\eta_{\alpha \prim}}{\sqrt{2 \epsilon_{\alpha \prim}}} \right)
  - \delta_{\alpha \beta}
  \frac{\eta_{\alpha \prim}}{2\epsilon_{\alpha \prim} W_{\prim}}
  \frac{\partial Z_{\alpha \com}}{\partial \phi_{\gamma \prim}}
  - \delta_{\alpha \gamma}
  \frac{\eta_{\alpha \prim}}{2 \epsilon_{\alpha \prim} W_{\prim}}
  \frac{\partial Z_{\alpha \com}}{\partial \phi_{\beta \prim}} \\
  \mbox{} +
  \frac{1}{W_{\prim} \sqrt{2 \epsilon_{\alpha \prim}}}
  \frac{\partial^2 Z_{\alpha \com}}{\partial \phi_{\beta \prim}
  \partial \phi_{\gamma \prim}} ,
  \label{Nthree}
\end{eqnarray}
where $\delta_{\alpha \beta \gamma}$ is unity when 
$\alpha =\beta = \gamma$ and is zero otherwise.

Finally, we must relate the derivatives of $Z_{\alpha \com}$ to 
the slow-roll parameters. After differentiating Eq.~\eref{Z} twice with 
respect to the field values on $\prim$ and making use of 
Eqs.~\eref{E} and~\eref{epsa}--\eref{etaa}, we find that
\begin{equation}
  \frac{\partial Z_{\alpha \com}}{\partial \phi_{\beta \prim}}
  = \frac{\sqrt{2} W_{\prim}}{\sqrt{\epsilon_{\beta \prim}}}
  \A_{\alpha\beta} ,
  \label{Adef}
\end{equation}
\begin{equation}
  \frac{\partial^2 Z_{\alpha \com}}{\partial \phi_{\gamma \prim}
  \partial \phi_{\beta \prim}} = - \delta_{\beta\gamma}
  \frac{\eta_{\beta \prim}}{\sqrt{2\epsilon_{\beta\prim}}}
  \frac{\partial Z_{\alpha \com}}{\partial \phi_{\beta \prim}} +
  \frac{2 W_{\prim}}{\sqrt{\epsilon_{\beta \prim}
  \epsilon_{\gamma \prim}}}
  \A_{\alpha\beta\gamma} ,
\end{equation}
where we have defined the totally symmetric matrices $\A_{\alpha\beta}$
\cite{Vernizzi:2006ve,Battefeld:2006sz} and $\A_{\alpha\beta\gamma}$
such that
\begin{equation}
  \A_{\alpha\beta} = - \frac{W_{\com}^2}{W_{\prim}^2}
  \sum_{\delta} Y_\delta E_{\alpha\delta}
  E_{\beta\delta} ,
  \label{Aaa}
\end{equation}
and 
\begin{eqnarray}
  \fl\nonumber
  \A_{\alpha\beta\gamma} = \frac{W_{\com}^3}{W_{\prim}^3}
  \sum_{\delta} \Bigg\{ X_{\delta} E_{\alpha\delta} E_{\beta\delta}
  E_{\gamma\delta} \\
  \mbox{} + \sum_{\kappa} Y_{\kappa}(Y_{\delta} - \epsilon_{\delta \com})
  \left[ E_{\alpha \kappa} E_{\beta \delta} E_{\gamma \delta}
       + E_{\beta \kappa} E_{\gamma \delta} E_{\alpha \delta}
       + E_{\gamma \kappa} E_{\alpha \delta} E_{\beta \delta} \right]
  \Bigg\} .
  \label{Aaaa}
\end{eqnarray}

Since $N_{, \alpha\beta\prim}$ and $N_{, \alpha\beta\gamma\prim}$
are derived from the partial differentiation of a scalar quantity,
they are symmetric under the exchanges
$\alpha \leftrightharpoons \beta$
and $\alpha \leftrightharpoons \beta \leftrightharpoons \gamma$,
respectively. Indeed, this symmetry is manifest since 
$\A_{\alpha\beta}$ and $\A_{\alpha\beta\gamma}$ are explicitly 
symmetric under an interchange of all indices. 

\section{An analytic formula for $\tnl$}
\label{sec:tnl}

Eqs.~\eref{None}--\eref{Nthree} 
represent expressions for the derivatives of $N$ given entirely in terms 
of the values of the potentials 
and slow-roll parameters \eref{epsilonsro}--\eref{xisro} 
evaluated on $\com$ and $\prim$.
It is now possible to derive an analytic expression 
for the non-linearity parameter $\tnl$ in terms of these 
parameters by evaluating the corresponding expressions 
for the quantities $\Delta\tnl^{(2)}$, $\Delta\tnl^{(3)}$ and
$\Delta\tnl^{(4)}$ defined in Eqs.~\eref{tnlb}--\eref{tnld}.
We begin by writing
\begin{equation}
  \Delta\tnl =
    \frac{1}{\left(
    \sum_\alpha \frac{u_\alpha^2}{\epsilon_{\alpha\prim}} \right)^3}
    \sum_r t^{(r)} ,
  \label{total-tnl}
\end{equation}
where $r \in \{ 1, 2, 3, 4 \}$.
Eq.~\eref{tnla} implies that $t^{(1)}$ is given by 
\begin{equation}
  t^{(1)} =
  \frac{2\Ffact_4}{\Tfact}
  \Big( \sum_{\alpha}
    \frac{u_{\alpha}^2}{\epsilon_{\alpha \prim}} \Big)^2
  \label{tfacta}
\end{equation}
and, by employing Eqs.~\eref{None}, \eref{Ntwo}--\eref{Nthree}, \eref{Aaa} and
\eref{Aaaa}, we find that the remaining $t^{(r)}$ are given by 
\begin{equation}
  \fl
  \frac{1}{\Kfact} t^{(2)} =
  \Big( \sum_{\alpha} \frac{u_{\alpha}^2}{\epsilon_{\alpha\prim}}
  \Big) \left[
  \sum_{\alpha} u_{\alpha}
  \left(1 - \frac{\eta_{\alpha\prim}}{2\epsilon_{\alpha\prim}} u_\alpha
  \right)
  +
  \sum_{\alpha\beta}
  \frac{u_{\alpha}}{\epsilon_{\alpha \prim}} \A_{\alpha\beta}
  \right]
  ,
  \label{tfactb}
\end{equation}
\begin{eqnarray}
  \fl\nonumber
  \frac{1}{4}t^{(3)} =
  \sum_{\alpha} \frac{u_{\alpha}^2}{\epsilon_{\alpha \prim}} \left(
    1 - \frac{\eta_{\alpha \prim}}{2\epsilon_{\alpha \prim}} u_{\alpha}
  \right)^2 +
  \sum_{\alpha\beta} 2 \frac{u_{\alpha} u_{\beta}}
    {\epsilon_{\alpha \prim} \epsilon_{\beta \prim}}
    \left(1 - \frac{\eta_{\alpha \prim}}{2\epsilon_{\alpha\prim}} u_{\alpha}
    \right) \A_{\alpha\beta} \\ \mbox{} +
  \sum_{\alpha\beta\gamma} \frac{u_{\beta} u_{\gamma}}
    {\epsilon_{\alpha \prim} \epsilon_{\beta \prim} \epsilon_{\gamma \prim}}
    \A_{\alpha\beta} \A_{\alpha\gamma} ,
  \label{tfactc}
\end{eqnarray}
and
\begin{eqnarray}
  \fl\nonumber
  \frac{\Tfact}{8 \sum_i k_i^3} t^{(4)} =
  \sum_{\alpha} \frac{u_{\alpha}^3}{2\epsilon_{\alpha \prim}^2}
  \left( \frac{u_{\alpha}}{\epsilon_{\alpha\prim}}
    (\eta_{\alpha \prim}^2 - \xi_{\alpha \prim}) - \eta_{\alpha \prim}
  \right)
  - \sum_{\alpha\beta} \frac{3}{2} \frac{u_{\alpha}^2 u_{\beta}}
    {\epsilon_{\alpha \prim}^2 \epsilon_{\beta \prim}}
    \eta_{\alpha \prim}
    \A_{\alpha\beta} \\ \mbox{} +
  \sum_{\alpha\beta\gamma}
    \frac{u_{\alpha} u_{\beta} u_{\gamma}}
      {\epsilon_{\alpha \prim}\epsilon_{\beta \prim} \epsilon_{\gamma\prim}}
    \A_{\alpha\beta\gamma} ,
  \label{tfactd}
\end{eqnarray}
respectively. 

Eqs.~\eref{total-tnl}--\eref{tfactd} are valid at any time during 
slow-roll inflation and represent 
the complete, tree-level non-gaussianity imprinted in the trispectrum 
for an inflationary scenario driven by multiple, uncoupled fields. 

\section{The horizon-crossing approximation}
\label{sec:horizoncross}

In this Section, we compute the non-gaussianity in $\tnl$
for the class of inflationary models where 
the potential $W(\phi)$ is given by a sum of monomials:
\begin{equation}
  W(\phi) = \sum_{\alpha} \lambda_{\alpha} \phi_{\alpha}^{\Gamma_{\alpha}} .
  \label{Nflation}
\end{equation}
This class includes the Nflation
\cite{Dimopoulos:2005ac,Easther:2005zr,Kim:2006ys}
and assisted inflationary  
\cite{Liddle:1998jc,Malik:1998gy,Copeland:1999cs,Kanti:1999vt,Green:1999vv}
scenarios. Nflation \cite{Dimopoulos:2005ac} 
is a realization of assisted quadratic 
inflation within the context of string theory compactifications. 
The scalar fields that drive inflation arise
as axionic degrees of freedom 
associated with shift symmetries in the compact manifold.
Although the axion potentials are sinusoidal, they can be approximated by
a quadratic form $\frac{1}{2} \lambda_{\alpha} \phi^2_{\alpha}$ 
when the fields are in the vicinity of their minima. 

In many inflationary models, such as Nflation and 
assisted inflation, the scalar fields settle into their 
minima during the final stages of accelerated expansion. In this case, 
the end-point (on the final surface $\com$) 
of the integral~\eref{efolds} is unimportant. 
This implies that the number of e-folds, $N$, becomes independent 
of the field values $\phi_{\alpha \com}$. Moreover, 
since the parameter $Z_{\alpha \com}$ quantifies 
the effect of including this end-point on the evolution of
the curvature perturbation, it too should be negligible.
Ignoring such contributions is known as 
the ``horizon-crossing'' approximation and is equivalent to  
specifying $Z_{\alpha \com} \approx \A_{\alpha\beta}
\approx \A_{\alpha\beta\gamma} \approx 0$ in Eqs.~\eref{tfactb}--\eref{tfactd}.
This implies that all quantities in the final expression for $\zeta$ 
are evaluated at the time of horizon exit on the surface $\prim$. 

The horizon-crossing approximation has been employed 
previously to argue that the 
non-gaussianity imprinted in the bispectrum is always small in multi-field 
inflation \cite{Alabidi:2005qi,Seery:2005gb,Kim:2006te}. 
Alabidi \& Lyth \cite{Alabidi:2005qi} have also 
used it to estimate the contribution $t^{(3)}$ to 
the non-linearity parameter $\tnl$. However, 
they did not employ the full tree-level formulae
given by Eqs. \eref{tfacta}--\eref{tfactd}. We will include all these 
contributions in what follows and 
assume that the horizon-crossing
approximation is valid. We will further assume that 
the coupling constants, $\lambda_{\alpha}$, and powers, $\Gamma_{\alpha}$, 
in the potential \eref{Nflation} are arbitrary, subject only 
to the condition that a slow-roll phase of inflation 
can be realised. It then follows that the number of e-folds from the 
surface $\prim$ to $\com$ is given by
\begin{equation}
  \label{Nprim}
  N=\frac{1}{2} \sum_{\alpha}  \frac{\phi^2_{\alpha \prim}}{\Gamma_{\alpha}} .
\end{equation}

We begin by considering the contribution from $t^{(4)}$. 
Since we are assuming that $\A_{\alpha\beta} \approx \A_{\alpha\beta\gamma} 
\approx 0$, it follows from Eq. \eref{Nprim} that $t^{(4)} = 0$. 
Hence, this term does not contribute to the non-linearity parameter 
$\tnl$. On the other hand, 
the remaining terms \eref{tfacta}--\eref{tfactc} yield
\begin{equation}
  \tnl = \frac{r}{8} \frac{\Ffact_4}{\Tfact} +
  \frac{r^2}{256} \left( \Kfact \sum_{\alpha}
    \frac{\lambda_{\alpha}}{W_{\prim} \Gamma_{\alpha}}
    \phi_{\alpha \prim}^{\Gamma_{\alpha}} +
    \frac{r}{2} \sum_{\alpha} \frac{\phi_{\alpha \prim}^2}{\Gamma_{\alpha}^3}
  \right) ,
  \label{tnl-monomial}
\end{equation}
where 
\begin{equation}
\label{defr}
  r = \frac{8 \powerspectrum_{\prim}}{\powerspectrum_{\zeta}} =
  16 \left( \sum_{\alpha} \frac{u_{\alpha}^2}{\epsilon_{\alpha \prim}}
  \right)^{-1} 
  = 8 \left( \sum_{\alpha} \frac{\phi^2_{\alpha \prim}}{\Gamma^2_{\alpha}} 
  \right)^{-1}  
\end{equation}
is the tensor-to-scalar ratio and 
$\powerspectrum_{\zeta} = \powerspectrum_{\prim}
\sum_{\alpha} N_{,\alpha \prim}^2$ is the power spectrum of $\zeta$. 
Eq. (\ref{tnl-monomial}) depends explicitly on the couplings, $\lambda_{\alpha}$,
and the initial conditions, $\phi_{\alpha \prim}$, 
which are expressed through the field values 
on the initial slice $\prim$. Hence, 
the non-gaussianity measured by the trispectrum is model-dependent. 

Further conclusions may be drawn if the powers 
$\Gamma_{\alpha}$ in the potential \eref{Nflation} 
take the same value, $\Gamma_{\alpha} \equiv \Gamma$, as in
Nflation for example. 
In this case, the tensor-to-scalar ratio is  
directly related to the number of e-folds such that $r=4\Gamma /N$
\cite{Alabidi:2005qi}.
As a result, Eq.~\eref{tnl-monomial} may be further 
simplified and we find that 
\begin{equation}
  \tnl = \frac{r}{8} \frac{\Ffact_4}{\Tfact} +
  \frac{r^2}{256 \Gamma} ( \Kfact + 4 ) .
  \label{nflation}
\end{equation}
Note that any dependence on the coupling parameters  
and the initial conditions of the fields 
has now disappeared.

The strongest upper bound 
on $r$ is imposed by the WMAP data \cite{Spergel:2006hy}, which implies that 
$r < 0.55$ at 95\% confidence. 
The contribution for $t^{(1)}$
is known to give a contribution to
$|\Delta\tnl|$ that is bounded from above by $0.06r$ \cite{Seery:2006vu}.
Even assuming that
$r$ saturates the WMAP upper bound, this gives no more 
than $|\Delta\tnl^{(1)} | \sim 1/30$, which is too small 
to be observable. The contribution from the other terms 
in Eq. (\ref{nflation}) will be maximized 
when $\Kfact$ is maximized and  
this is expected to occur on equilateral configurations. In this case we find 
that $\Kfact \approx -1.7$. Hence, the WMAP bound on $r$  
implies that in the equilateral limit
the second term in Eq.~\eref{nflation} can not exceed 
$(3 \times 10^{-3})/\Gamma$ and this is also unobservably small.  
Although it is likely 
that some memory of the initial conditions will remain 
in the isocurvature-driven
contributions $\A_{\alpha\beta}$ and $\A_{\alpha\beta\gamma}$, the
considerations of Ref.~\cite{Battefeld:2006sz} suggest that these too 
will be small, with the most significant 
contribution arising when the fields
have a narrow spectrum of masses (or couplings). 
We conclude, therefore, that the non-gaussianity from assisted 
inflationary scenarios based on monomial potentials with equal powers 
is unlikely to be visible in the trispectrum.

\section{Conclusions}
\label{sec:conclude}

In this paper, we have employed the $\delta N$ formalism
to derive the first complete estimate at tree-level for
the trispectrum non-linearity parameter $\tnl$  
in uncoupled, multi-field inflation. Our result assumes 
that the slow-roll approximation is valid
between horizon exit, and the time of interest, which we have taken
to be before the end of inflation.
We have considered an arbitrary number of fields
and accounted for the influence of isocurvature perturbations 
in generating non-gaussianities on superhorizon scales. 

We have employed our expression within the context of the 
horizon-crossing approximation to compute the
non-gaussianity imprinted in the trispectrum
by an epoch of inflation driven by a sum of arbitrary, monomial potentials
with arbitrary coupling constants. In general, it was found that 
the non-linearity parameter $\tnl$ is model-dependent, 
since it is sensitive to the initial conditions and 
the values of the coupling parameters.  
However, when the powers in the monomials 
take the same value, the non-linearity parameter 
becomes independent of the initial conditions and
couplings. In this class of models, we are able to estimate an upper 
bound on $\tnl$ which is determined by the tensor-to-scalar ratio.  
We find that the non-gaussianity is always small and will not
be observable to upcoming CMB experiments, such as the Planck
satellite. Alternative scenarios exist in which the non-gaussianity
may be larger, such as the curvaton, for which a comparable
prediction for the trispectrum exists \cite{Sasaki:2006kq}.
A future measurement of the parameter $\tnl$ offers the possibility of
a clean experimental comparison between these scenarios.

\textit{Note added:} After this paper was submitted, a similar
work \cite{Byrnes:2006vq} appeared
in which the authors use the $\delta N$ formalism to
compute an expression for $\tnl$
comparable to Eqs.~\eref{tnlb}--\eref{tnld},
and give an explicit slow-roll expression 
comparable to Eqs.~\eref{tfactc}--\eref{tfactd}
in the single-field case, when the bispectrum is zero and the
horizon-crossing approximation applies.
We have verified that in the appropriate special cases
the results of Ref. \cite{Byrnes:2006vq} exactly agree with those
presented in this paper.

\ack
DS is supported by PPARC grant PPA/G/S/2003/00076. DS would like to thank
F. Vernizzi and D. Wands for useful conversations
and correspondence. We thank D. Lyth and the
Department of Physics at Lancaster University for their hospitality
during the workshop \emph{Non-gaussianity from Inflation}, June 2006.

\section*{References}

\providecommand{\href}[2]{#2}\begingroup\raggedright\endgroup


\begin{thebibliography}{10}

\bibitem{Spergel:2006hy}
D.~N. Spergel {\em et~al.}, {\it Wilkinson microwave anisotropy probe ({WMAP})
  three year results: Implications for cosmology},
  \href{http://xxx.lanl.gov/abs/astro-ph/0603449}{{\tt astro-ph/0603449}}.

\bibitem{Martin:2006rs}
J.~Martin and C.~Ringeval, {\it Inflation after {WMAP3}: Confronting the
  slow-roll and exact power spectra to {CMB} data},  {\sl JCAP} {\bf 0608}
  (2006) 009, [\href{http://xxx.lanl.gov/abs/astro-ph/0605367}{{\tt
  astro-ph/0605367}}].

\bibitem{Kinney:2006qm}
W.~H. Kinney, E.~W. Kolb, A.~Melchiorri, and A.~Riotto, {\it Inflation model
  constraints from the {Wilkinson} microwave anisotropy probe three-year data},
   {\sl Phys. Rev.} {\bf D74} (2006) 023502,
  [\href{http://xxx.lanl.gov/abs/astro-ph/0605338}{{\tt astro-ph/0605338}}].

\bibitem{Starobinsky:1980te}
A.~A. Starobinsky, {\it A new type of isotropic cosmological models without
  singularity},  {\sl Phys. Lett.} {\bf B91} (1980) 99--102.

\bibitem{Sato:1980yn}
K.~Sato, {\it First order phase transition of a vacuum and expansion of the
  universe},  {\sl Mon. Not. Roy. Astron. Soc.} {\bf 195} (1981) 467--479.

\bibitem{Guth:1980zm}
A.~H. Guth, {\it The inflationary universe: A possible solution to the horizon
  and flatness problems},  {\sl Phys. Rev.} {\bf D23} (1981) 347--356.

\bibitem{Hawking:1981fz}
S.~W. Hawking and I.~G. Moss, {\it Supercooled phase transitions in the very
  early universe},  {\sl Phys. Lett.} {\bf B110} (1982) 35.

\bibitem{Albrecht:1982wi}
A.~Albrecht and P.~J. Steinhardt, {\it Cosmology for grand unified theories
  with radiatively induced symmetry breaking},  {\sl Phys. Rev. Lett.} {\bf 48}
  (1982) 1220--1223.

\bibitem{Linde:1981mu}
A.~D. Linde, {\it A new inflationary universe scenario: A possible solution of
  the horizon, flatness, homogeneity, isotropy and primordial monopole
  problems},  {\sl Phys. Lett.} {\bf B108} (1982) 389--393.

\bibitem{Linde:1983gd}
A.~D. Linde, {\it Chaotic inflation},  {\sl Phys. Lett.} {\bf B129} (1983)
  177--181.

\bibitem{Bardeen:1983qw}
J.~M. Bardeen, P.~J. Steinhardt, and M.~S. Turner, {\it Spontaneous creation of
  almost scale-free density perturbations in an inflationary universe},  {\sl
  Phys. Rev.} {\bf D28} (1983) 679.

\bibitem{Guth:1982ec}
A.~H. Guth and S.~Y. Pi, {\it Fluctuations in the new inflationary universe},
  {\sl Phys. Rev. Lett.} {\bf 49} (1982) 1110--1113.

\bibitem{Hawking:1982cz}
S.~W. Hawking, {\it The development of irregularities in a single bubble
  inflationary universe},  {\sl Phys. Lett.} {\bf B115} (1982) 295.

\bibitem{Hawking:1982my}
S.~W. Hawking and I.~G. Moss, {\it Fluctuations in the inflationary universe},
  {\sl Nucl. Phys.} {\bf B224} (1983) 180.

\bibitem{Bartolo:2004if}
N.~Bartolo, E.~Komatsu, S.~Matarrese, and A.~Riotto, {\it Non-gaussianity from
  inflation: Theory and observations},  {\sl Phys. Rept.} {\bf 402} (2004)
  103--266, [\href{http://xxx.lanl.gov/abs/astro-ph/0406398}{{\tt
  astro-ph/0406398}}].

\bibitem{Rigopoulos:2004gr}
G.~I. Rigopoulos and E.~P.~S. Shellard, {\it Non-linear inflationary
  perturbations},  {\sl JCAP} {\bf 0510} (2005) 006,
  [\href{http://xxx.lanl.gov/abs/astro-ph/0405185}{{\tt astro-ph/0405185}}].

\bibitem{Rigopoulos:2004ba}
G.~I. Rigopoulos, E.~P.~S. Shellard, and B.~W. van Tent, {\it A simple route to
  non-gaussianity in inflation},  {\sl Phys. Rev.} {\bf D72} (2005) 083507,
  [\href{http://xxx.lanl.gov/abs/astro-ph/0410486}{{\tt astro-ph/0410486}}].

\bibitem{Rigopoulos:2005xx}
G.~I. Rigopoulos, E.~P.~S. Shellard, and B.~W. van Tent, {\it Non-linear
  perturbations in multiple-field inflation},  {\sl Phys. Rev.} {\bf D73}
  (2006) 083521, [\href{http://xxx.lanl.gov/abs/astro-ph/0504508}{{\tt
  astro-ph/0504508}}].

\bibitem{Allen:2005ye}
L.~E. Allen, S.~Gupta, and D.~Wands, {\it Non-gaussian perturbations from
  multi-field inflation},  {\sl JCAP} {\bf 0601} (2006) 006,
  [\href{http://xxx.lanl.gov/abs/astro-ph/0509719}{{\tt astro-ph/0509719}}].

\bibitem{Maldacena:2002vr}
J.~M. Maldacena, {\it Non-gaussian features of primordial fluctuations in
  single field inflationary models},  {\sl JHEP} {\bf 05} (2003) 013,
  [\href{http://xxx.lanl.gov/abs/astro-ph/0210603}{{\tt astro-ph/0210603}}].

\bibitem{Komatsu:2001rj}
E.~Komatsu and D.~N. Spergel, {\it Acoustic signatures in the primary microwave
  background bispectrum},  {\sl Phys. Rev.} {\bf D63} (2001) 063002,
  [\href{http://xxx.lanl.gov/abs/astro-ph/0005036}{{\tt astro-ph/0005036}}].

\bibitem{Verde:1999ij}
L.~Verde, L.-M. Wang, A.~Heavens, and M.~Kamionkowski, {\it Large-scale
  structure, the cosmic microwave background, and primordial non-gaussianity},
  {\sl Mon. Not. Roy. Astron. Soc.} {\bf 313} (2000) L141--L147,
  [\href{http://xxx.lanl.gov/abs/astro-ph/9906301}{{\tt astro-ph/9906301}}].

\bibitem{Okamoto:2002ik}
T.~Okamoto and W.~Hu, {\it The angular trispectra of {CMB} temperature and
  polarization},  {\sl Phys. Rev.} {\bf D66} (2002) 063008,
  [\href{http://xxx.lanl.gov/abs/astro-ph/0206155}{{\tt astro-ph/0206155}}].

\bibitem{Kogo:2006kh}
N.~Kogo and E.~Komatsu, {\it Angular trispectrum of {CMB} temperature
  anisotropy from primordial non-gaussianity with the full radiation transfer
  function},  {\sl Phys. Rev.} {\bf D73} (2006) 083007,
  [\href{http://xxx.lanl.gov/abs/astro-ph/0602099}{{\tt astro-ph/0602099}}].

\bibitem{Boubekeur:2005fj}
L.~Boubekeur and D.~H. Lyth, {\it Detecting a small perturbation through its
  non- gaussianity},  {\sl Phys. Rev.} {\bf D73} (2006) 021301,
  [\href{http://xxx.lanl.gov/abs/astro-ph/0504046}{{\tt astro-ph/0504046}}].

\bibitem{Lyth:2005fi}
D.~H. Lyth and Y.~Rodr\'{\i}guez, {\it The inflationary prediction for
  primordial non- gaussianity},  {\sl Phys. Rev. Lett.} {\bf 95} (2005) 121302,
  [\href{http://xxx.lanl.gov/abs/astro-ph/0504045}{{\tt astro-ph/0504045}}].

\bibitem{Alabidi:2005qi}
L.~Alabidi and D.~H. Lyth, {\it Inflation models and observation},  {\sl JCAP}
  {\bf 0605} (2006) 016, [\href{http://xxx.lanl.gov/abs/astro-ph/0510441}{{\tt
  astro-ph/0510441}}].

\bibitem{Creminelli:2005hu}
P.~Creminelli, A.~Nicolis, L.~Senatore, M.~Tegmark, and M.~Zaldarriaga, {\it
  Limits on non-gaussianities from {WMAP} data},  {\sl JCAP} {\bf 0605} (2006)
  004, [\href{http://xxx.lanl.gov/abs/astro-ph/0509029}{{\tt
  astro-ph/0509029}}].

\bibitem{Creminelli:2006rz}
P.~Creminelli, L.~Senatore, M.~Zaldarriaga, and M.~Tegmark, {\it Limits on
  $f_nl$ parameters from wmap 3yr data},
  \href{http://xxx.lanl.gov/abs/astro-ph/0610600}{{\tt astro-ph/0610600}}.

\bibitem{Cooray:2006km}
A.~Cooray, {\it 21-cm background anisotropies can discern primordial
  non-gaussianity from slow-roll inflation},
  \href{http://xxx.lanl.gov/abs/astro-ph/0610257}{{\tt astro-ph/0610257}}.

\bibitem{Lyth:2006qz}
D.~H. Lyth and D.~Seery, {\it Classicality of the primordial perturbations},
  \href{http://xxx.lanl.gov/abs/astro-ph/0607647}{{\tt astro-ph/0607647}}.

\bibitem{Seery:2006vu}
D.~Seery, J.~E. Lidsey, and M.~S. Sloth, {\it The inflationary trispectrum},
  \href{http://xxx.lanl.gov/abs/astro-ph/0610210}{{\tt astro-ph/0610210}}.

\bibitem{Seery:2005wm}
D.~Seery and J.~E. Lidsey, {\it Primordial non-gaussianities in single field
  inflation},  {\sl JCAP} {\bf 0506} (2005) 003,
  [\href{http://xxx.lanl.gov/abs/astro-ph/0503692}{{\tt astro-ph/0503692}}].

\bibitem{Chen:2006nt}
X.~Chen, M.-X. Huang, S.~Kachru, and G.~Shiu, {\it Observational signatures and
  non-gaussianities of general single field inflation},
  \href{http://xxx.lanl.gov/abs/hep-th/0605045}{{\tt hep-th/0605045}}.

\bibitem{Huang:2006eh}
M.-X. Huang and G.~Shiu, {\it The inflationary trispectrum for models with
  large non- gaussianities},
  \href{http://xxx.lanl.gov/abs/hep-th/0610235}{{\tt hep-th/0610235}}.

\bibitem{Lyth:2004gb}
D.~H. Lyth, K.~A. Malik, and M.~Sasaki, {\it A general proof of the
  conservation of the curvature perturbation},  {\sl JCAP} {\bf 0505} (2005)
  004, [\href{http://xxx.lanl.gov/abs/astro-ph/0411220}{{\tt
  astro-ph/0411220}}].

\bibitem{Malik:2005cy}
K.~A. Malik, {\it Gauge-invariant perturbations at second order: Multiple
  scalar fields on large scales},  {\sl JCAP} {\bf 0511} (2005) 005,
  [\href{http://xxx.lanl.gov/abs/astro-ph/0506532}{{\tt astro-ph/0506532}}].

\bibitem{Malik:2006ir}
K.~A. Malik, {\it A not so short note on the {K}lein-{G}ordon equation at
  second order},  \href{http://xxx.lanl.gov/abs/astro-ph/0610864}{{\tt
  astro-ph/0610864}}.

\bibitem{Lyth:2005du}
D.~H. Lyth and Y.~Rodr\'{\i}guez, {\it Non-gaussianity from the second-order
  cosmological perturbation},  {\sl Phys. Rev.} {\bf D71} (2005) 123508,
  [\href{http://xxx.lanl.gov/abs/astro-ph/0502578}{{\tt astro-ph/0502578}}].

\bibitem{Seery:2005gb}
D.~Seery and J.~E. Lidsey, {\it Primordial non-gaussianities from
  multiple-field inflation},  {\sl JCAP} {\bf 0509} (2005) 011,
  [\href{http://xxx.lanl.gov/abs/astro-ph/0506056}{{\tt astro-ph/0506056}}].

\bibitem{Lyth:2005qj}
D.~H. Lyth and I.~Zaballa, {\it A bound concerning primordial non-gaussianity},
   {\sl JCAP} {\bf 0510} (2005) 005,
  [\href{http://xxx.lanl.gov/abs/astro-ph/0507608}{{\tt astro-ph/0507608}}].

\bibitem{Zaballa:2006pv}
I.~Zaballa, Y.~Rodr\'{\i}guez, and D.~H. Lyth, {\it Higher order contributions
  to the primordial non- gaussianity},  {\sl JCAP} {\bf 0606} (2006) 013,
  [\href{http://xxx.lanl.gov/abs/astro-ph/0603534}{{\tt astro-ph/0603534}}].

\bibitem{Vernizzi:2006ve}
F.~Vernizzi and D.~Wands, {\it Non-gaussianities in two-field inflation},  {\sl
  JCAP} {\bf 0605} (2006) 019,
  [\href{http://xxx.lanl.gov/abs/astro-ph/0603799}{{\tt astro-ph/0603799}}].

\bibitem{Kim:2006te}
S.~A. Kim and A.~R. Liddle, {\it Nflation: non-gaussianity in the
  horizon-crossing approximation},
  \href{http://xxx.lanl.gov/abs/astro-ph/0608186}{{\tt astro-ph/0608186}}.

\bibitem{Alabidi:2006hg}
L.~Alabidi, {\it Non-gaussianity for a two component hybrid model of
  inflation},  \href{http://xxx.lanl.gov/abs/astro-ph/0604611}{{\tt
  astro-ph/0604611}}.

\bibitem{Enqvist:2004bk}
K.~Enqvist and A.~Vaihkonen, {\it Non-gaussian perturbations in hybrid
  inflation},  {\sl JCAP} {\bf 0409} (2004) 006,
  [\href{http://xxx.lanl.gov/abs/hep-ph/0405103}{{\tt hep-ph/0405103}}].

\bibitem{Piao:2006nm}
Y.-S. Piao, {\it On perturbation spectra of {N}-flation},
  \href{http://xxx.lanl.gov/abs/gr-qc/0606034}{{\tt gr-qc/0606034}}.

\bibitem{Gupta:2002kn}
S.~Gupta, A.~Berera, A.~F. Heavens, and S.~Matarrese, {\it Non-gaussian
  signatures in the cosmic background radiation from warm inflation},  {\sl
  Phys. Rev.} {\bf D66} (2002) 043510,
  [\href{http://xxx.lanl.gov/abs/astro-ph/0205152}{{\tt astro-ph/0205152}}].

\bibitem{Gupta:2005nh}
S.~Gupta, {\it Dynamics and non-gaussianity in the weak-dissipative warm
  inflation scenario},  {\sl Phys. Rev.} {\bf D73} (2006) 083514,
  [\href{http://xxx.lanl.gov/abs/astro-ph/0509676}{{\tt astro-ph/0509676}}].

\bibitem{Lyth:2002my}
D.~H. Lyth, C.~Ungarelli, and D.~Wands, {\it The primordial density
  perturbation in the curvaton scenario},  {\sl Phys. Rev.} {\bf D67} (2003)
  023503, [\href{http://xxx.lanl.gov/abs/astro-ph/0208055}{{\tt
  astro-ph/0208055}}].

\bibitem{Lyth:2006gd}
D.~H. Lyth, {\it Non-gaussianity and cosmic uncertainty in curvaton-type
  models},  {\sl JCAP} {\bf 0606} (2006) 015,
  [\href{http://xxx.lanl.gov/abs/astro-ph/0602285}{{\tt astro-ph/0602285}}].

\bibitem{Malik:2006pm}
K.~A. Malik and D.~H. Lyth, {\it A numerical study of non-gaussianity in the
  curvaton scenario},  \href{http://xxx.lanl.gov/abs/astro-ph/0604387}{{\tt
  astro-ph/0604387}}.

\bibitem{Sasaki:2006kq}
M.~Sasaki, J.~V{\"{a}}liviita, and D.~Wands, {\it Non-gaussianity of the
  primordial perturbation in the curvaton model},
  \href{http://xxx.lanl.gov/abs/astro-ph/0607627}{{\tt astro-ph/0607627}}.

\bibitem{Enqvist:2005pg}
K.~Enqvist and S.~Nurmi, {\it Non-gaussianity in curvaton models with nearly
  quadratic potential},  {\sl JCAP} {\bf 0510} (2005) 013,
  [\href{http://xxx.lanl.gov/abs/astro-ph/0508573}{{\tt astro-ph/0508573}}].

\bibitem{Enqvist:2004ey}
K.~Enqvist, A.~Jokinen, A.~Mazumdar, T.~Multamaki, and A.~Vaihkonen, {\it
  Non-gaussianity from preheating},  {\sl Phys. Rev. Lett.} {\bf 94} (2005)
  161301, [\href{http://xxx.lanl.gov/abs/astro-ph/0411394}{{\tt
  astro-ph/0411394}}].

\bibitem{Barnaby:2006cq}
N.~Barnaby and J.~M. Cline, {\it Nongaussian and nonscale-invariant
  perturbations from tachyonic preheating in hybrid inflation},  {\sl Phys.
  Rev.} {\bf D73} (2006) 106012,
  [\href{http://xxx.lanl.gov/abs/astro-ph/0601481}{{\tt astro-ph/0601481}}].

\bibitem{Battefeld:2006sz}
T.~Battefeld and R.~Easther, {\it Non-gaussianities in multi-field inflation},
  \href{http://xxx.lanl.gov/abs/astro-ph/0610296}{{\tt astro-ph/0610296}}.

\bibitem{Huang:2005nd}
Q.-G. Huang and K.~Ke, {\it Non-gaussianity in {KKLMMT} model},  {\sl Phys.
  Lett.} {\bf B633} (2006) 447--452,
  [\href{http://xxx.lanl.gov/abs/hep-th/0504137}{{\tt hep-th/0504137}}].

\bibitem{Starobinsky:1986fx}
A.~A. Starobinsky, {\it Multicomponent de sitter (inflationary) stages and the
  generation of perturbations},  {\sl JETP Lett.} {\bf 42} (1985) 152--155.

\bibitem{Sasaki:1995aw}
M.~Sasaki and E.~D. Stewart, {\it A general analytic formula for the spectral
  index of the density perturbations produced during inflation},  {\sl Prog.
  Theor. Phys.} {\bf 95} (1996) 71--78,
  [\href{http://xxx.lanl.gov/abs/astro-ph/9507001}{{\tt astro-ph/9507001}}].

\bibitem{Langlois:2006vv}
D.~Langlois and F.~Vernizzi, {\it Nonlinear perturbations of cosmological
  scalar fields},  \href{http://xxx.lanl.gov/abs/astro-ph/0610064}{{\tt
  astro-ph/0610064}}.

\bibitem{Wands:2000dp}
D.~Wands, K.~A. Malik, D.~H. Lyth, and A.~R. Liddle, {\it A new approach to the
  evolution of cosmological perturbations on large scales},  {\sl Phys. Rev.}
  {\bf D62} (2000) 043527,
  [\href{http://xxx.lanl.gov/abs/astro-ph/0003278}{{\tt astro-ph/0003278}}].

\bibitem{Rigopoulos:2003ak}
G.~I. Rigopoulos and E.~P.~S. Shellard, {\it The separate universe approach and
  the evolution of nonlinear superhorizon cosmological perturbations},  {\sl
  Phys. Rev.} {\bf D68} (2003) 123518,
  [\href{http://xxx.lanl.gov/abs/astro-ph/0306620}{{\tt astro-ph/0306620}}].

\bibitem{Salopek:1990jq}
D.~S. Salopek and J.~R. Bond, {\it Nonlinear evolution of long wavelength
  metric fluctuations in inflationary models},  {\sl Phys. Rev.} {\bf D42}
  (1990) 3936--3962.

\bibitem{Sloth:2006az}
M.~S. Sloth, {\it On the one loop corrections to inflation and the {CMB}
  anisotropies},  {\sl Nucl. Phys.} {\bf B748} (2006) 149--169,
  [\href{http://xxx.lanl.gov/abs/astro-ph/0604488}{{\tt astro-ph/0604488}}].

\bibitem{Lyth:1998xn}
D.~H. Lyth and A.~Riotto, {\it Particle physics models of inflation and the
  cosmological density perturbation},  {\sl Phys. Rept.} {\bf 314} (1999)
  1--146, [\href{http://xxx.lanl.gov/abs/hep-ph/9807278}{{\tt
  hep-ph/9807278}}].

\bibitem{Dimopoulos:2005ac}
S.~Dimopoulos, S.~Kachru, J.~McGreevy, and J.~G. Wacker, {\it N-flation},
  \href{http://xxx.lanl.gov/abs/hep-th/0507205}{{\tt hep-th/0507205}}.

\bibitem{Easther:2005zr}
R.~Easther and L.~McAllister, {\it Random matrices and the spectrum of
  {N}-flation},  {\sl JCAP} {\bf 0605} (2006) 018,
  [\href{http://xxx.lanl.gov/abs/hep-th/0512102}{{\tt hep-th/0512102}}].

\bibitem{Kim:2006ys}
S.~A. Kim and A.~R. Liddle, {\it Nflation: Multi-field inflationary dynamics
  and perturbations},  {\sl Phys. Rev.} {\bf D74} (2006) 023513,
  [\href{http://xxx.lanl.gov/abs/astro-ph/0605604}{{\tt astro-ph/0605604}}].

\bibitem{Liddle:1998jc}
A.~R. Liddle, A.~Mazumdar, and F.~E. Schunck, {\it Assisted inflation},  {\sl
  Phys. Rev.} {\bf D58} (1998) 061301,
  [\href{http://xxx.lanl.gov/abs/astro-ph/9804177}{{\tt astro-ph/9804177}}].

\bibitem{Malik:1998gy}
K.~A. Malik and D.~Wands, {\it Dynamics of assisted inflation},  {\sl Phys.
  Rev.} {\bf D59} (1999) 123501,
  [\href{http://xxx.lanl.gov/abs/astro-ph/9812204}{{\tt astro-ph/9812204}}].

\bibitem{Copeland:1999cs}
E.~J. Copeland, A.~Mazumdar, and N.~J. Nunes, {\it Generalized assisted
  inflation},  {\sl Phys. Rev.} {\bf D60} (1999) 083506,
  [\href{http://xxx.lanl.gov/abs/astro-ph/9904309}{{\tt astro-ph/9904309}}].

\bibitem{Kanti:1999vt}
P.~Kanti and K.~A. Olive, {\it On the realization of assisted inflation},  {\sl
  Phys. Rev.} {\bf D60} (1999) 043502,
  [\href{http://xxx.lanl.gov/abs/hep-ph/9903524}{{\tt hep-ph/9903524}}].

\bibitem{Green:1999vv}
A.~M. Green and J.~E. Lidsey, {\it Assisted dynamics of multi-scalar field
  cosmologies},  {\sl Phys. Rev.} {\bf D61} (2000) 067301,
  [\href{http://xxx.lanl.gov/abs/astro-ph/9907223}{{\tt astro-ph/9907223}}].

\bibitem{Byrnes:2006vq}
C.~T. Byrnes, M.~Sasaki, and D.~Wands, {\it The primordial trispectrum from
  inflation},  \href{http://xxx.lanl.gov/abs/astro-ph/0611075}{{\tt
  astro-ph/0611075}}.

\end{thebibliography}
\end{document}